\documentclass[pra,twocolumn,showpacs,amsfonts,amsmath,floatfix]{revtex4}

\usepackage[pdftex]{graphicx}
\usepackage{array}
\usepackage{ulem}

\begin{document}
\title{30-Hz relative linewidth watt output power 1.65~$\mu$m continuous-wave singly resonant optical parametric oscillator}

\author{Aliou Ly$^{1}$}
\author{Christophe Siour$^{1}$ }
\author{Fabien Bretenaker$^{1}$}
\email{Fabien.Bretenaker@u-psud.fr}
 \affiliation {$^1$Laboratoire Aim\'e Cotton, Universit\'e Paris-Sud, ENS Paris-Saclay, CNRS, Universit\'e Paris-Saclay, Orsay, France
 }


\begin{abstract}
We built a 1-watt cw singly resonant optical parametric oscillator operating at an idler wavelength of 1.65~$\mu$m for application to quantum interfaces. The non resonant idler is frequency stabilized by side-fringe locking on a relatively high-finesse Fabry-Perot cavity, and the influence of intensity noise is carefully analyzed. A relative linewidth down to the sub-kHz level (about 30\,Hz over 2\,s) is achieved. A very good long term stability  is obtained for both frequency and intensity.  
\end{abstract}
\maketitle

\section{Introduction}
\label{intro}
Building efficient quantum communication networks requires the implementation of quantum light-matter interfaces in the communication channels in order to overcome two major limitations. First, the attenuation of light in telecom C-band optical fibers significantly limits the reachable communication distances to few tens of kilometers at the single photon level. Second, the most efficient single photon detectors such as Si avalanche photodiodes lie into the visible range of the spectrum. Quantum memories for photons have then been proposed to increase the communication distances by allowing the synchronization of quantum relays. But it appears that the most efficient memories do not operate at telecom wavelengths \cite{Bussieres2013}.

To efficiently interface a memory to a telecom single photon source and take advantage of Si APDs, one can use quantum frequency conversion via nonlinear processes in a $\chi^{(2)}$ medium \cite{Kumar1990}, which has been shown to be relatively efficient and to preserve the quantum state of the photon \cite{Tanzilli2005}. Since then, numerous experimental realizations involving single photon up-conversion and using  sum-frequency generation in a $\chi^{\mathrm{(2)}}$ medium have been performed. For instance, a recent example is the demonstration of storage of up-converted telecom photons into a quantum memory for visible photons \cite{Maring2014}. Cheng et $\it{al}.$ have up-converted near-IR photons to the visible with practically no background photons \cite{Cheng2015}. Concerning non-classical Gaussian states, Vollmer et $\it{al}.$ succeeded in up-converting squeezed vacuum states from 1550 to 532~nm \cite{Vollmer2014}. 

In most of the experiments described above, single photon up-conversion is performed in periodically-poled lithium niobate (PPLN). Then one problem that arises is the noise that may exist at the up-converted wavelength even in the absence of single photons to be converted. A large part of this noise is due to anti-Stokes stimulated Raman scattering (SRS) in lithium niobate \cite{Pelc2011}. To overcome this limitation, several strategies were explored. Some authors have tried to perform an efficient spectral filtering of the up-converted signal \cite{Kuo2013}, while others have tried a cascaded single photon up-conversion \cite{Pelc2012}. But the most promising approach would be to pump at a wavelength much longer than that of the signal to be up-converted. Moreover, the availability of a broadly tunable pump would certainly make it easier to minimize the SRS noise for a given signal or up-converted wavelength. Moreover, although the frequency noise of the pump will not significantly affect the efficiency of the quantum interface \cite{Maring2014}, it might reduce the quantum memory efficiency depending on the used protocol. Up to now, the pump laser sources that have been used  to up-convert near-infrared single photons are CW laser diodes \cite{Cheng2015}, external-cavity diode lasers followed by a tapered amplifier \cite{Maring2014}, thulium fiber lasers amplified by a thulium doped fiber amplifier \cite{Shentu2013}, or a monolithic optical parametric oscillator (OPO) combined with a fiber amplifier \cite{Pelc2011}. 

However, recently, it has been shown that the non resonant wave in a singly-resonant OPO (SRO) could be stabilized by transferring the pump noise to the resonant wave \cite{Ly2015,Ricciardi2015,Silander2015}. This permits to take advantage of the efficient power transfer from the pump to the non resonant wave while stabilizing the frequency noise below the pump frequency noise. Consequently, the aim of the present paper is to take advantage of this source architecture to build a stabilized SRO operating around 1.65 $\mu$m. We choose this wavelength as a test wavelength for up-converting  photons at telecom wavelengths (typically around 1535 nm) to the wavelength of rubidium quantum memories (795 nm). In the second section, we describe the SRO we have built. The third section then reports the power, tunability, and mode quality of this SRO. Finally, Sec. 4 is devoted to a study of the power and frequency stability of the OPO, with a comparison of free running and frequency locked operations.

\section{Description of the OPO}
\label{setup}
\begin{figure}
\centering\includegraphics[width=1.0\columnwidth]{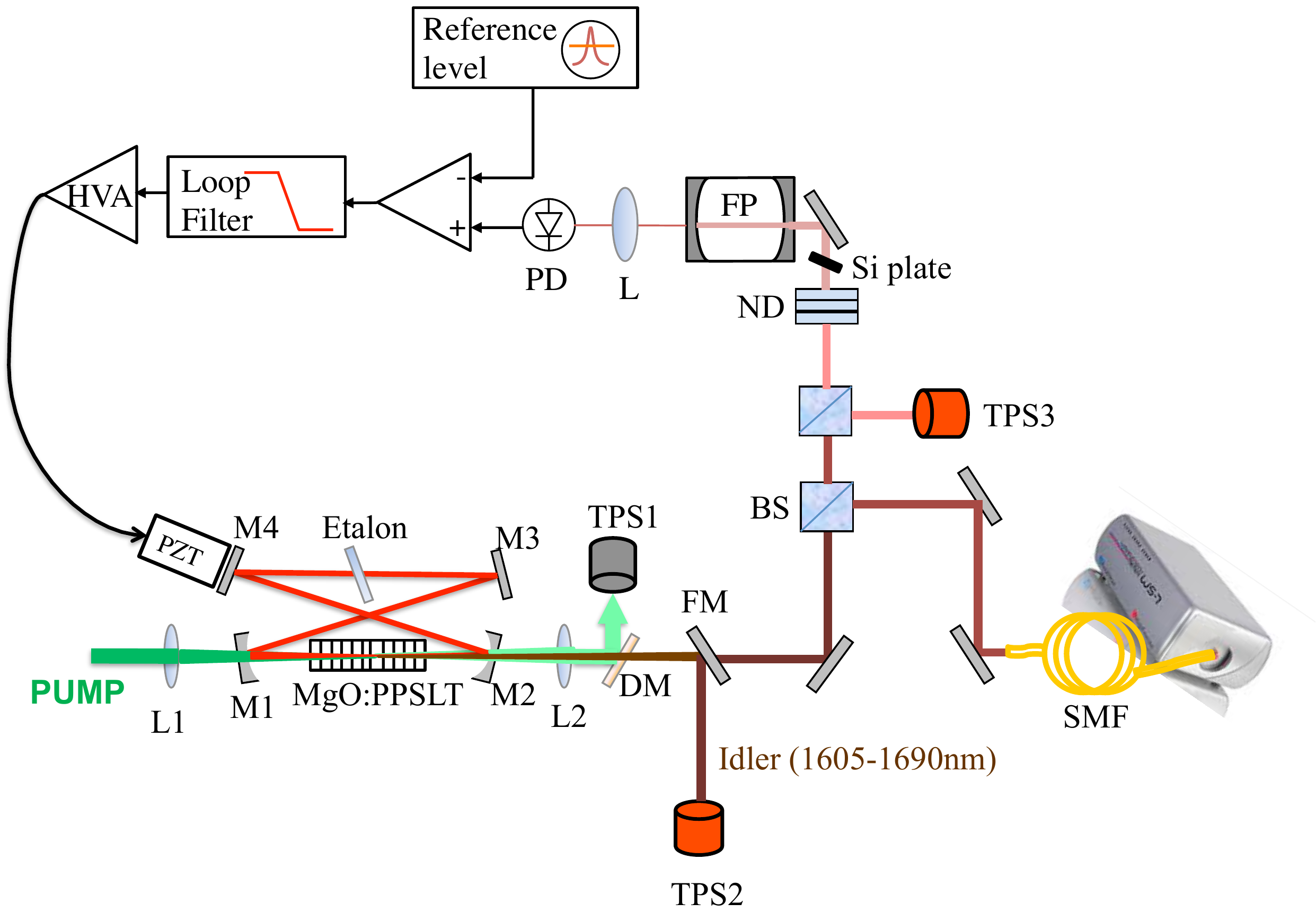}
\vspace*{0cm} 
\caption{Schematic of the experimental setup. HVA: high-voltage amplifier. PZT: piezoelectric transducer. PD: Photodiode. L: focusing lens. FP: Fabry-Perot cavity. DM: Dichroic Mirror. BS: Beamsplitter Cube. ND: Neutral Density . TPS: Thermal Power Sensor. FM: Flippable Mirror. SMF: Single-Mode Fiber. MgO:PPSLT: MgO-doped periodically poled stochiometric lithium tantalate crystal.}
\label{Fig011} 
\end{figure}

The experimental setup is depicted in Fig.\,\ref{Fig011}. The SRO is pumped at 532~nm by a cw 10~W single-frequency Coherent Verdi  laser and is based on a 30-mm long MgO-doped periodically poled stoichiometric lithium tantalate (PPSLT) crystal ($d_{\mathrm{eff}}\simeq11~\mathrm{pm/V}$) manufactured and coated by HC Photonics. This crystal contains a single grating with a period of 8.61~$\mu$m. Its faces are anti-reflection coated (reflection coefficient smaller than 0.5~\%) for the pump, signal, idler wavelengths and are wedged by 1\,$^{\circ}$  to avoid \'etalon effects. It is designed to lead to quasi-phase matching conditions for a signal wavelength in the 776-796~nm range, depending on the temperature. The OPO cavity is a 1.095-m long ring cavity and consists in four mirrors. The two concave mirrors both have a 150~mm radius of curvature. The two other mirrors are planar. All mirrors are designed to exhibit a high reflectivity $(R\geq~99.9~\%)$ between 776~nm and 796~nm and a transmission larger than 95~\% at 532~nm and between 1605~nm and 1690~nm. This allows  the OPO to be singly resonant, the resonant wave being the signal. The calculated waist of the signal beam at the middle of the PPSLT crystal is 43~$\mu$m. The pump beam is focused by a 160~mm focal-length lens (L1) to a 35~$\mu$m radius waist inside the PPSLT crystal. Finally, to ensure stable single-frequency operation of the OPO, an uncoated YAG \'etalon  is inserted in the second waist of the cavity. Different \'etalon thicknesses were tried, leading all to stable single-frequency operation of the OPO. In the following, the different results that are reproduced have been obtained either with a 150~$\mu$m thick or 250~$\mu$m thick \'etalon.

\section{SRO performances}
\begin{figure}[h!]
\centering\includegraphics[width=1.0\columnwidth]{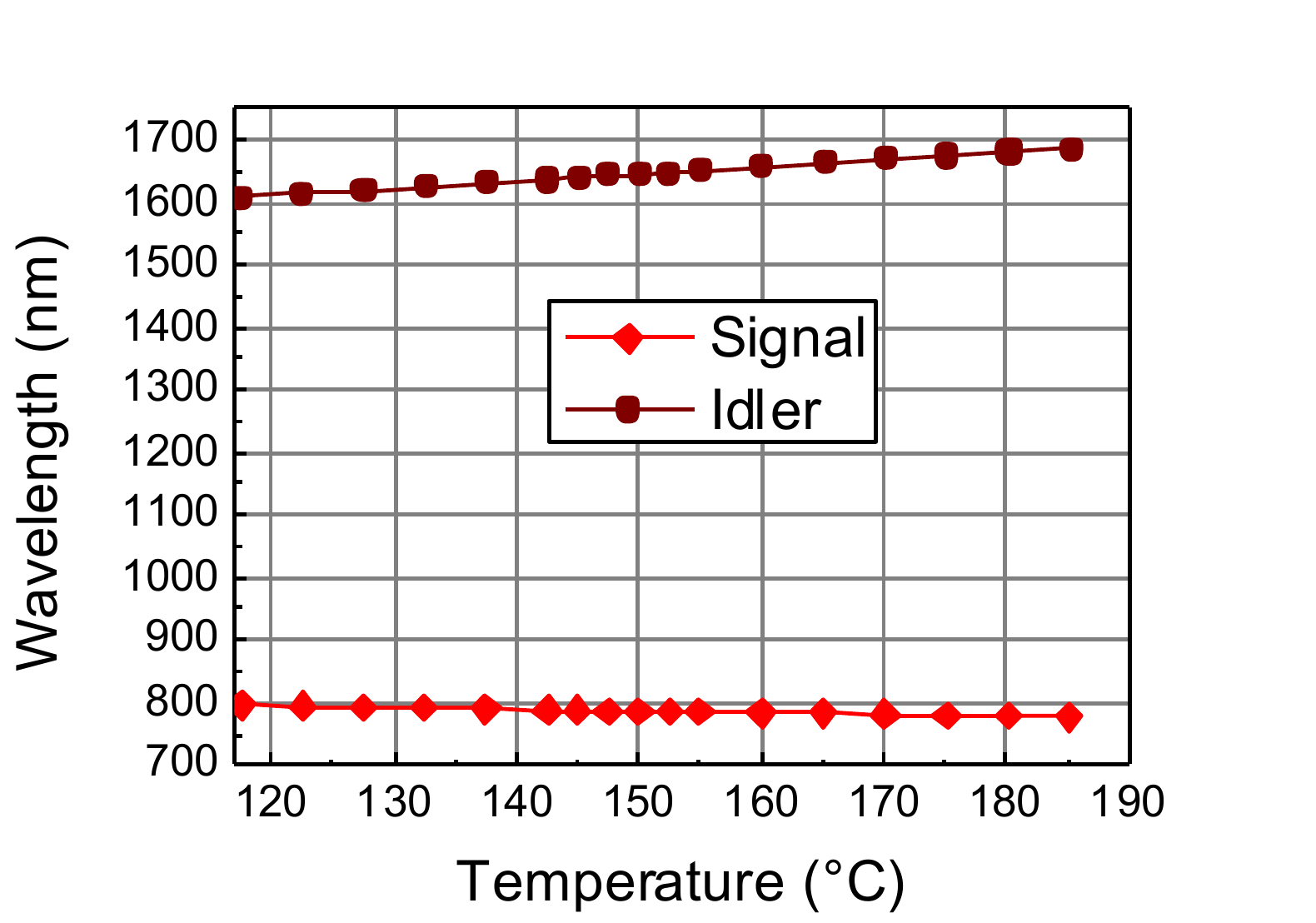}
\vspace*{0cm} 
\caption{Output signal and idler wavelengths versus PPSLT crystal temperature. The full lines are present just to guide the eyes.}
\label{Fig022} 
\end{figure}

We first investigate the wavelength tuning characteristics of the OPO at a fixed pump power. For this measurement we use a spectrometer (AvaSpec 2048-2) to monitor the signal wavelength. The idler wavelength is then calculated based on energy conservation. The crystal is heated by a homemade oven and its temperature is automatically controlled to 0.1\,$^{\circ}$C by a LFI-3751 Thermoelectric Temperature Controller and monitored by a thermistance. The measured thermal tuning curves are reported in Fig.\,\ref{Fig022}. One can see that the idler wavelength can be tuned from 1607.3~nm to 1686.09~nm (777.235~nm to 795.205~nm for the signal) when the PPSLT temperature is varied from 117.5\,$^{\circ}$C to 185.2\,$^{\circ}$C. The experimental slopes for the tunability of the OPO were estimated to be of 1.16~nm/$^{\circ}$C for the idler and -0.26~nm/$^{\circ}$C for the signal, in reasonable agreement with the dispersion curves available in the literature \cite{Bruner2003}.

\begin{figure}
\centering\includegraphics[width=1.0\columnwidth]{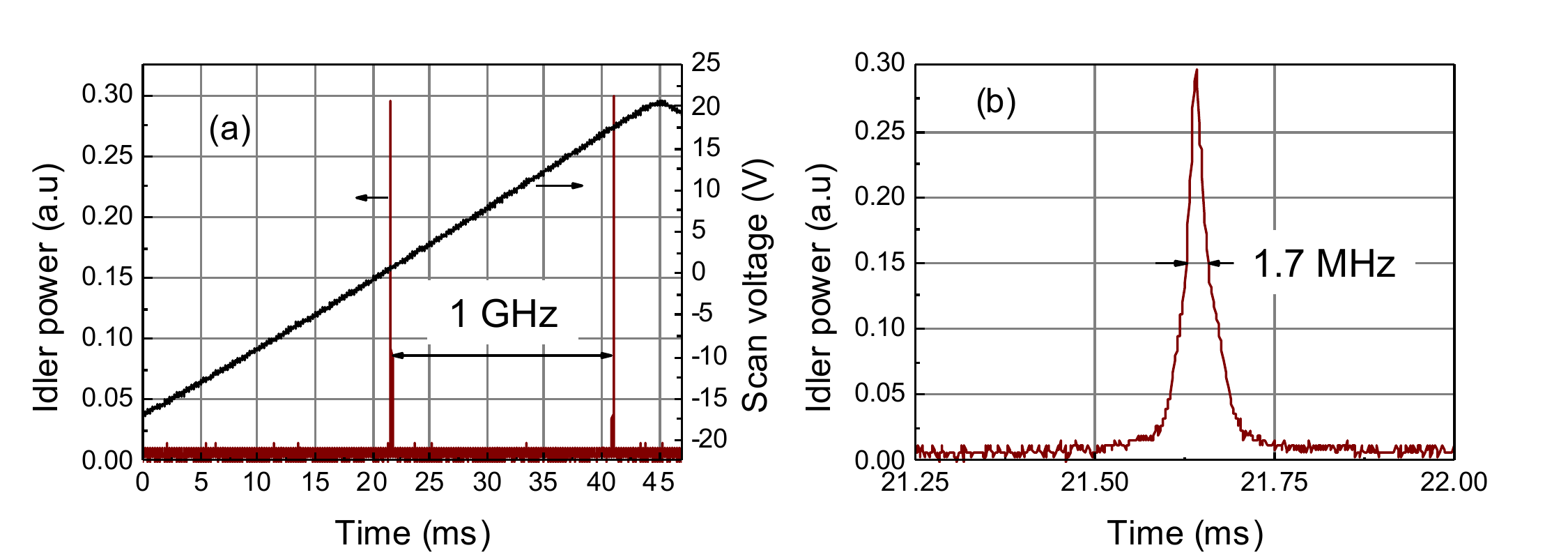}
 \vspace*{0cm} 
\caption{ (a) Observation of the idler spectrum using a scanning Fabry-Perot interferometer with a 1~GHz free spectral range. (b) Zoom on one of the peaks. The peak linewidth is limited by the Fabry-Perot finesse.}
\label{Fig033} 
\end{figure}

In the following, we maintain the temperature of the PPSLT crystal at T=152.5\,$^{\circ}$C. This corresponds to a signal wavelength equal to 786~nm and an idler wavelength of 1646~nm. The OPO output beams are collimated with a  200\,mm focal-length lens (L2). At this temperature, we check thanks to a $\Delta=1\,\mathrm{GHz}$ free spectral range Fabry-Perot interferometer operating at the idler wavelength that the OPO stably oscillates on a single longitudinal mode. The corresponding signals can be seen in Fig.\,\ref{Fig033}, with a 150-$\mu$m-thick \'etalon. One can clearly see in Fig.\,\ref{Fig033}(a) that the OPO operates on a signal frequency. A zoom on one of the peaks shows that the peak full width at half maximum is equal to 1.74 MHz, which is actually limited by the finesse $F=580$ of the Fabry-Perot (FP).

\begin{figure}[h]
\centering\includegraphics[width=1.0\columnwidth]{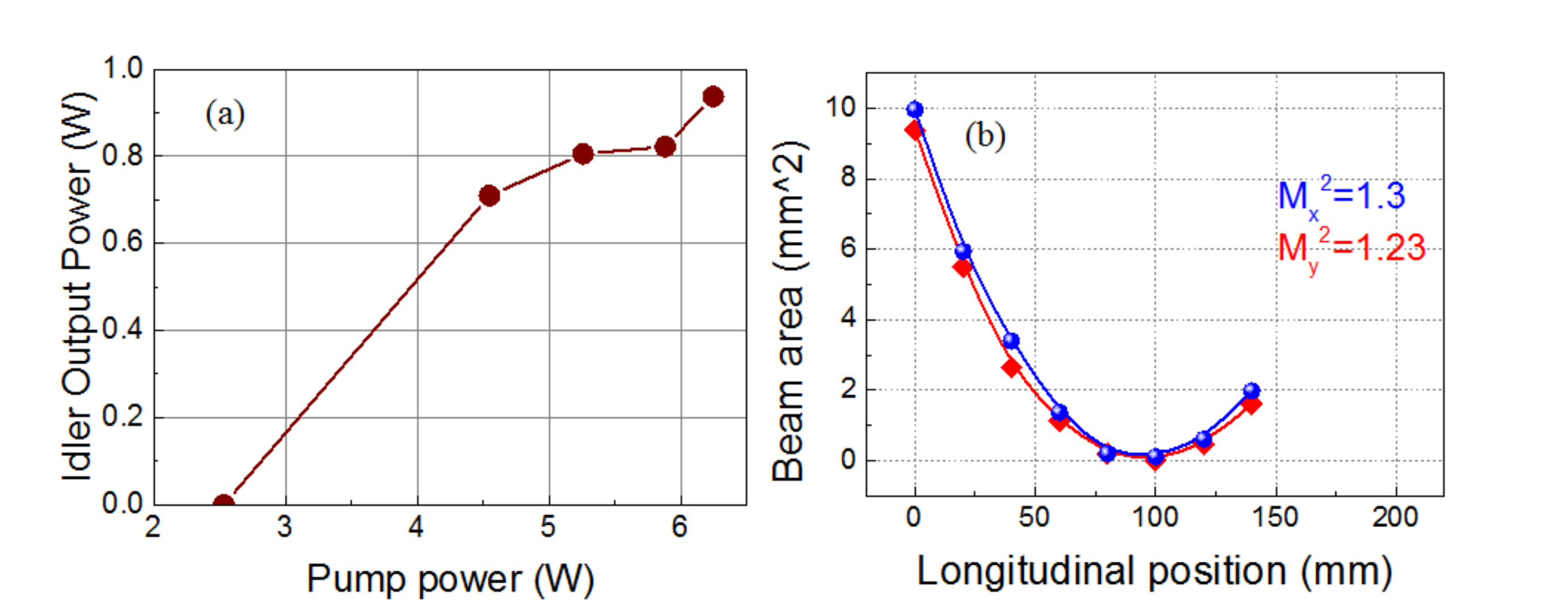}
 \vspace*{0cm} 
\caption{ (a) Measured evolution of the idler output power versus pump power. (b) Evolution of the square of the beam radius versus propagation distance in the two transverse dimensions $x$ (resp. $y$) corresponds to the direction orthogonal to (resp. inside) the plane of the cavity. The full lines correspond to polynomial fits and the symbols to the experimental data.}
\label{Fig044} 
\end{figure}

We then measure the idler output power when the pump power is gradually decreased. Figure \ref{Fig044}(a) displays the obtained results. The plotted idler power corresponds to the one just at the output of mirror M2 of the OPO cavity. The pump power is measured  just before the input concave mirror M1 of the cavity. We can see that a watt level idler output power can be reached at 6.5~W pump power. At a pump power smaller than 4.5~W, the idler output power quickly drops. The OPO threshold is found to be equal to 2.5~W. This is comparable with the threshold pump power we calculate using a simple truncated Gaussian beam model \cite{Sutherland2003} with about 5~\% round-trip losses for the signal. This threshold value could of course be decreased by decreasing the losses. For example, in the absence of intracavity \'etalon, we have checked that the threshold decreases below 1~W. However, since this would lead to lower conversion efficiencies at larger pump powers \cite{Breuning2011}, we keep these losses in the following. 

In order to check the spatial quality of the idler beam, we precisely measure its $M^2$ factor. To this aim, we measure the evolution of the beam radius when it is focused by a 150-mm focal length lens. Beam radii measurements are performed using the knife-edge method. From the experimental data we plot the square of the beam radius along the propagation axis and we perform a polynomial fit to recover the $M^2$ parameter in the two transverse directions, as shown in Fig.\,\ref{Fig044}(b). As one can see, the idler beam is nearly diffraction limited. The slight discrepancy with respect to a perfect TEM$_{00}$ mode is attributed to thermal effects in the PPSLT crystal.

\section{Spectral purity, stability, and servo-locking}
Let us now focus on the noise properties of the OPO since this is a critical parameter in many quantum optics experiments. We focus on both the short-term spectral purity of the OPO (frequency noise, linewidth) and on its long term wavelength and power stability. In the following, all experimental results were obtained at a pump power of approximately 5~W before L1 and a crystal temperature fixed at 152.5\,$^{\circ}$C.

\subsection{Spectral purity}
We aim here at frequency locking the non-resonant idler of the OPO on a reference cavity. Preceding studies have shown that an OPO can be quite efficiently frequency stabilized by locking either on the side of the cavity transmission fringe \cite{Mhibik2010, Ly2015} or on the top of the transmission fringe \cite{Mhibik2011,Andrieux2011}. The first solution is more straightforward but the second one has the advantage of being immune to intensity fluctuations. Here, we choose the first solution, for its simplicity, but we carefully analyze the influence of intensity noise on the frequency noise of the locked OPO. Before designing the frequency servo-loop, we first need to have an estimation of the OPO frequency noise. More precisely, we first need to have an idea of the bandwidth of the idler frequency noise.

\subsubsection{Free running OPO}
In order to monitor the idler frequency fluctuations, we use,  as a frequency to intensity converter, the cavity that was used to observe the OPO spectrum in Fig. \ref{Fig033}. The idler power incident on that reference FP cavity is equal to 500~$\mu$W. As sketched in Fig.\,\ref{Fig011}, a silicon plate is used to completely absorb the residual pump signal.
 
 As shown in Appendix A, when the OPO frequency lies close to the middle of the Fabry-Perot fringe, the OPO frequency fluctuations are linearly converted into fluctuations of the transmitted intensity. The voltage produced by the amplified photodetector that follows the FP can thus be converted into a frequency deviation $\delta\nu(t)$ of the OPO idler using the following equation:
\begin{equation}
\delta\nu(t)=\frac{\Delta}{F\;V_{pp}}\delta V(t)\equiv K\delta V(t)\ ,\label{eq01}
\end{equation}
where $\Delta$ holds for the free spectral range of the cavity, $V_{pp}$ is the voltage difference between the maximum and minimum transmission of the cavity, and $\delta V(t)$ is the photodetector voltage deviation with respect to the middle of the FP fringe. Equation (\ref{eq01}) is valid only when the bandwidth of the fluctuations is much smaller than the cavity bandwidth $\Delta/F$, which will be the case in the following.

\begin{figure}[h]
\centering\includegraphics[width=1.0\columnwidth]{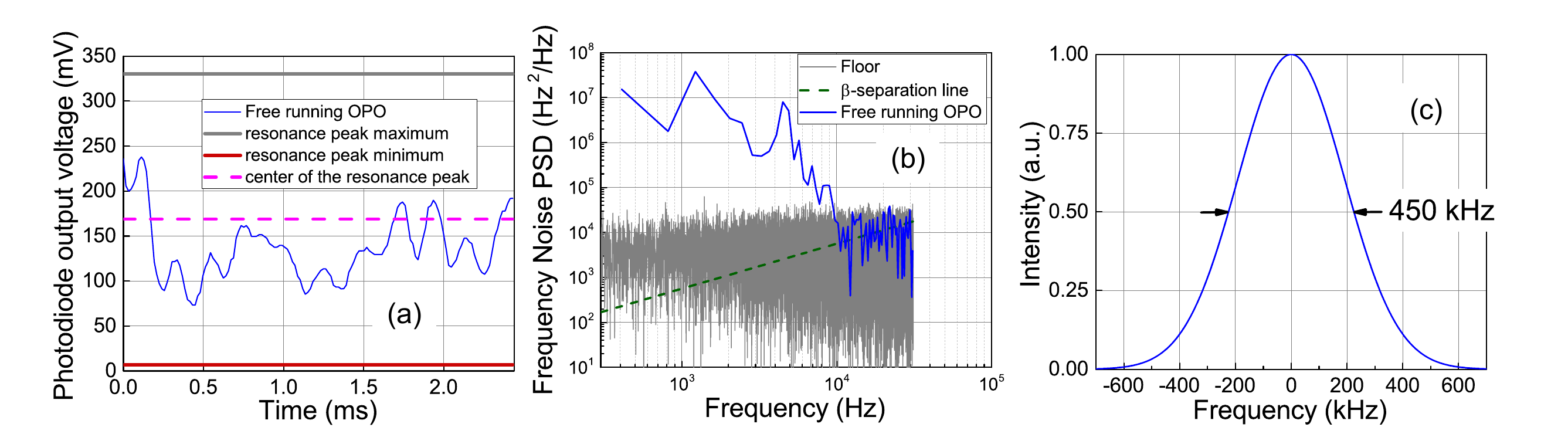}
 \vspace*{0cm} 
\caption{ Free running OPO frequency noise characteristics. (a) Blue line: time evolution of the FP transmission signal. The solid red and gray lines hold for the FP minimum and maximum transmissions, respectively. The dashed pink line is located at the center of the resonance peak. (b) Single-sided power spectral density $S_{\delta\nu}(f)$ of the idler frequency noise extracted from the blue signal in (a). The dashed olive line is the $\beta$ separation line. The gray line is the measurement noise floor. (c) Spectrum of the idler corresponding to the noise PSD in (b).}\label{Fig055} 
\end{figure}

We therefore tune the OPO idler frequency to the side of the resonance peak of the cavity, which has a linewidth of 1.7 MHz . Since the OPO frequency is not locked, it drifts quite fast and remains on the side of the FP fringe only for a short duration. Figure\,\ref{Fig055}(a) shows such a transmission signal in the free running mode for a  2.45-ms-long sample. Although the signal remains within the same side fringe of the cavity transmission, we can see that at some points it goes a bit far away from the middle of the fringe. This means that the linearization that has been performed to obtain Eq. (\ref{eq01}) is not strictly valid, leading to the fact that by using it we will slightly underestimate the frequency fluctuations. Nevertheless, this will provide us with a good order of magnitude for the OPO frequency noise. 

Following Eq. (\ref{eq01}) with $\Delta=1\,\mathrm{GHz}$, $F=580$, and $V_{pp}=320\,\mathrm{mV}$, we calculate the frequency fluctuations from the trace of Fig. \ref{Fig055}(a) with $K=5400\,\mathrm{Hz/mV}$. A first simple method to estimate the linewidth of the OPO is to extract the RMS value of these fluctuations, which we find to be $\Delta\nu_{\mathrm{RMS}}=190\,\mathrm{kHz}$. If we suppose for example that the laser spectrum is Gaussian, this leads to a FWHM given by\cite{Elliott1982} $\Delta\nu_{\mathrm{FWHM}}=2.35\;\Delta\nu_{\mathrm{RMS}}=450\,\mathrm{kHz}$.

We have also derived the power spectral density $S_{\delta\nu}(f)$ of the idler frequency noise by applying a fast Fourier transform to $\delta\nu(t)$ derived from the data of Fig.\,\ref{Fig055}(a). The result is reproduced in Fig.\,\ref{Fig055}(b). It can be clearly seen that the strongest components of the frequency noise spectrum lie at frequencies below 10 kHz. Thus, the use of a piezoelectric transducer carrying a cavity mirror should be fast enough to apply the relevant correction signal. The spectrum of the idler can be derived from the data of Fig.\,\ref{Fig055}(b) by first calculating the auto-correlation of the idler field $E(t)$ using the following expression \cite{Middleton1960,Elliott1982}:
\begin{equation}
R_E(\tau)\propto\exp\left[-\int_{-\infty}^{\infty}S_{\nu}(f)\frac{1-\cos (2\pi f \tau)}{f^2}\mathrm{d}f\right]\ ,\label{eq08}
\end{equation}
which, in our case, is limited to the frequency range $[f_1,f_2]=[408\,\mathrm{Hz}, 30\,\mathrm{kHz}]$. Using the Wiener-Khinchin theorem, we then compute the idler spectrum, reproduced in Fig.\,\ref{Fig055}(c). This spectrum indeed has a Gaussian shape, with a linewidth equal to 450~kHz for the $[f_1,f_2]$ bandwidth. Finally, we use the approach proposed by Di Domenico et $\it{al}$. \cite{DiDomenico2010} to plot the function $\beta(f)$ in Fig.\,\ref{Fig055}(b), defined as:
\begin{equation}
\beta(f)=\frac{8\ln(2)}{\pi^2}f\ ,\label{eqbeta}
\end{equation}
where $f$ is the Fourier frequency of the frequency noise PSD. We can see in Fig.\,\ref{Fig055}(b) that the part of the laser noise that is above $\beta(f)$, i. e., that contributes to the laser linewidth if we suppose that the lineshape is Gaussian, corresponds to instantaneous frequencies below 10~kHz, i. e., which are manageable by a simple piezoelectric transducer. Finally, if we use the expression \cite{DiDomenico2010}
\begin{equation}
\Delta\nu_{\mathrm{FWHM}}=\sqrt{8A\ln(2)}\ ,\label{eqDomenico}
\end{equation}
where
\begin{equation}
A=\int_{f_1}^{f_2}H\left[S_{\delta\nu}(f)-\beta(f)\right]S_{\delta\nu}(f)\mathrm{df}\ .\label{eqheavi}
\end{equation}
$H$ in Eq. (\ref{eqheavi}) is the Heaviside step function. Applying Eqs.\,(\ref{eqDomenico}) and (\ref{eqheavi}) to the data of in Fig.\,\ref{Fig055}(b) also gives a FWHM linewidth equal to 450~kHz.

\subsubsection{Locked OPO}
Since the results of the preceding section have shown that the bandwidth of the fluctuations is smaller than 10 kHz, and that the amplitude of these fluctuations over a few ms is in the MHz range, we can hope to correct them by applying a voltage to a piezoelectric transducer that carries one of the cavity mirrors (see Fig.\,\ref{Fig011}). The voltage produced by the photodetector that follows the FP reference is compared with a voltage that corresponds to the middle of the FP fringe. Then, the difference is sent to a PI controller whose output is amplified using a high-voltage amplifier before being applied to the piezoelectric transducer. By adjusting the gain of the loop and the corner frequency of the PI controller, we could lock the OPO idler frequency to the FP cavity. 

\begin{figure}[h]
\centering\includegraphics[width=1.0\columnwidth]{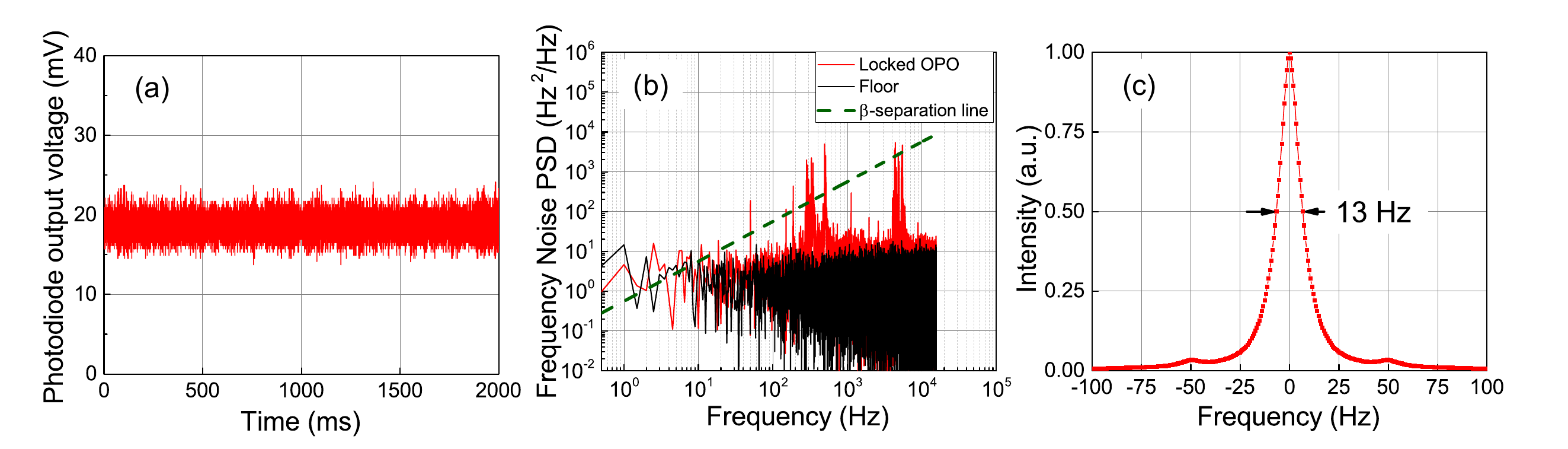}
 \vspace*{0cm} 
\caption{Locked OPO frequency noise characteristics. (a) Time evolution of the FP transmission of the non resonant idler. (b) Red line: single-sided power spectral density of the idler frequency fluctuations obtained from the signal in (a) \cite{Elliott1982,Middleton1960}. The dashed olive line is the $\beta$ separation line \cite{DiDomenico2010} and the black spectrum is the measurement noise floor. (c) Spectrum of the non resonant idler field deduced from the frequency noise PSD in (b).}\label{Fig066} 
\end{figure}
Figure\,\ref{Fig066}(a) shows the output voltage of the PD recorded  during 2\,s while the OPO is locked. Compared with the data recorded in the open loop regime, we have modified the gain of the photodiode preamplifier, leading to $V_{pp}=3.9\,\mathrm{V}$, leading to $K=440\,\mathrm{Hz/mV}$. We thus use Eq.\,(\ref{eq01}) to calculate the instantaneous relative frequency fluctuations of the idler with respect to the FP resonance, whose power spectral density is reproduced in Fig.\,\ref{Fig066}(b).

It can be clearly seen that, when the servo loop is closed, the relative frequency noise of the OPO idler is significantly reduced. The idler frequency noise PSD becomes quasi white in the considered frequency domain, with an average level of the order of a few $\mathrm{Hz}^2/\mathrm{Hz}$. The small peaks at lower frequencies (between 200 and 500 Hz) are reminiscent of the pump laser noise, not perfectly compensated by the servo-loop \cite{Mhibik2010}. The biggest noise peak located at about 5~kHz frequency reaches an amplitude equal to $10^3\,\mathrm{Hz}^2/\mathrm{Hz}$. It might be a resonance of the piezoelectric actuator since its frequency did not change when the parameters of the servo controller (gain and corner frequency) were modified.

The strong noise reduction when the loop is closed can be analyzed as follows. First, the PI loop filter, with a corner frequency set at 3\,kHz and a high static gain permits the correction of the noise over the entire idler frequency noise bandwidth, up to about $10\,\mathrm{kHz}$. Second, the decrease of the background noise level at high frequencies (above $10\,\mathrm{kHz}$) from a few $10^3\,\mathrm{Hz^2/Hz}$ in the free running case [see Fig.\,\ref{Fig055}(b)] to a few $10\,\mathrm{Hz^2/Hz}$ in the locked case [see Fig.\,\ref{Fig066}(b)] simply comes from the fact that the photodiode preamplifier gain has been changed, just changing the frequency fluctuation level equivalent of the detection noise.  

In the case of a quasi-white noise spectrum like the one of  Fig.\,\ref{Fig066}(b), we expect the laser lineshape to be no longer Gaussian but more Lorentzian. Following \cite{Elliott1982}, we thus estimate the laser linewidth using $\Delta\nu_{\mathrm{FWHM}}=\pi S_{\delta\nu}(f)$. By taking $S_{\delta\nu}\approx10\,\mathrm{Hz}^2/\mathrm{Hz}$, we obtain a linewidth of the order of  $30\,\mathrm{Hz}$. When we apply the $\beta$ line method to the frequency noise spectrum in Fig.\ \ref{Fig066}(b), we find  $\Delta\nu_{\mathrm{FWHM}}=14\,\mathrm{Hz}$ for the frequency interval $[0.5\;\mathrm{Hz},10\;\mathrm{Hz}]$. Finally, The computation of the idler spectrum in Fig.\ \ref{Fig066}(c) using Eq. (\ref{eq08}) leads to a FWHM linewidth of 13\,Hz over 2\,s, with a lineshape which is no longer Gaussian, as expected. Here also, like in the free running mode, the idler linewidth values found by these different methods are all in good agreement. One should not be surprised by the fact that the linewidth of the locked OPO is smaller than the RMS frequency noise value because the bandwidth considered here is about of 20\,kHz, which is much larger than the RMS noise (570\,Hz) \cite{Handbook2011}. Finally, one should notice that the peaks in the spectrum of Fig.\ \ref{Fig066}(b) do not contribute to the laser linewidth of Fig.\ \ref{Fig066}(c). Only the effect of the noise peak at $50\,\mathrm{Hz}$, coming from the power supply, is visible in Fig.\ \ref{Fig066}(c).

\subsection{Influence of the intensity noise on the spectral purity}
It is well known that a serious drawback of the side-of-fringe locking technique with respect to the Pound-Drever-Hall locking technique is that the former one is not immune to intensity fluctuations: any intensity fluctuation could be wrongly attributed to frequency fluctuations and lead the servo-lock to introduce extra frequency noise \cite{Mhibik2011}. It is thus of paramount importance to measure the intensity fluctuations and to quantify their influence on the frequency fluctuations of the locked OPO. To this aim, we recorded the idler intensity noise of the free running OPO. We recorded this intensity noise with the same detector as the one used after the reference Fabry-Perot cavity, and we attenuated the detected intensity in such a way that it is equal to the average intensity detected after the reference cavity when the OPO is locked. In order to evaluate the possible frequency error introduced by this intensity noise, we process the corresponding voltage using Eq. (\ref{eq01}), as if it were a frequency noise signal converted into an intensity noise. Figure \ref{Fig088} compares the spectrum of this intensity noise with the error signal noise when the OPO is locked.  We can see that the intensity noise of the free running OPO is slightly larger than the error signal noise when the OPO is locked. This means that the servo loop converts a small fraction of the OPO intensity noise into unwanted frequency fluctuations. In order to investigate the order of magnitude of this conversion, we write the photodiode voltage noise $\delta V(t)$ as the sum of two terms:
\begin{figure}[h!]
\centering\includegraphics[width=1.0\columnwidth]{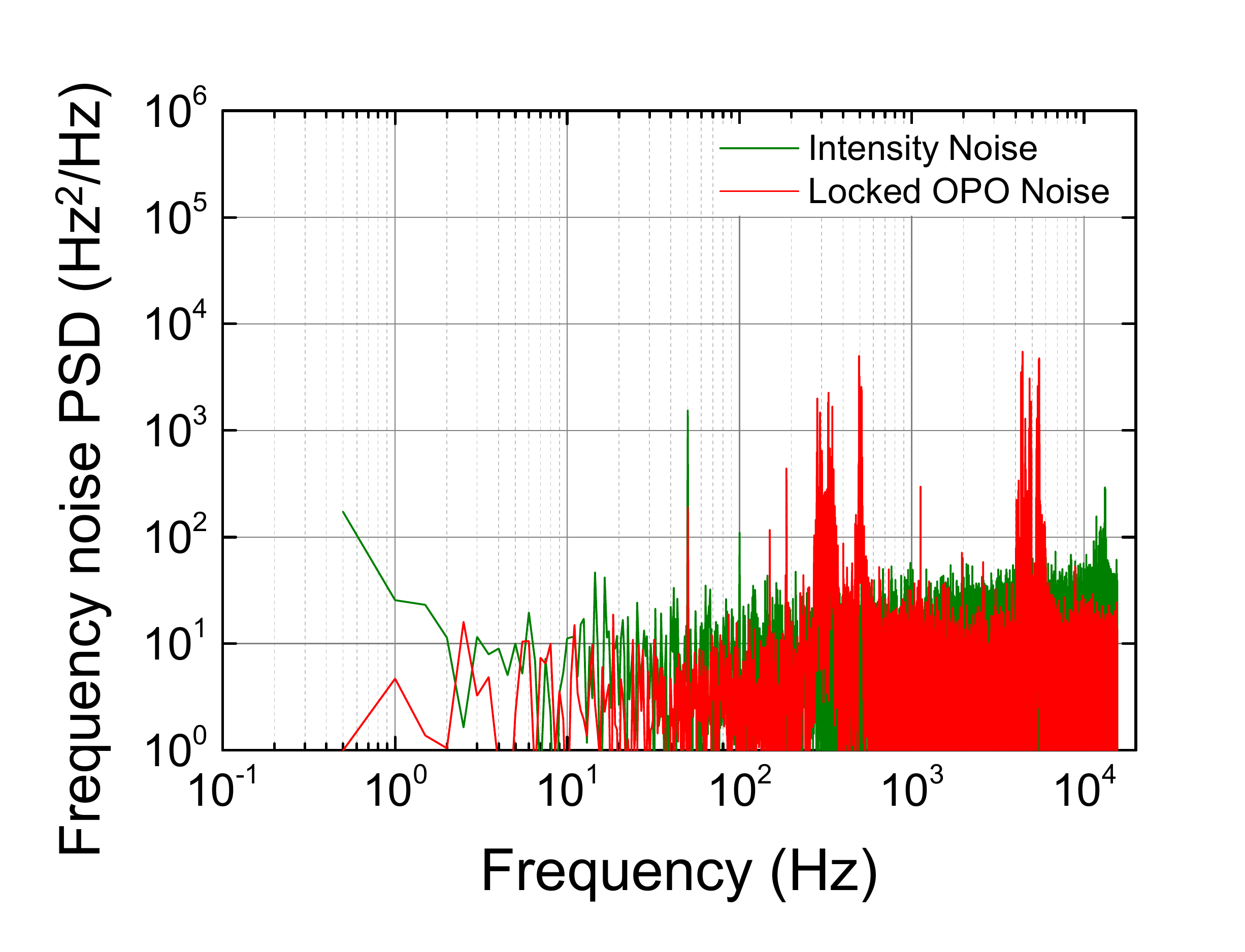}
  \vspace*{0cm} 
\caption{Single-sided power spectral density of the intensity noise of the non resonant idler, converted into frequency noise (see text), when the OPO is free running (green) and of the detected frequency error signal when the OPO is locked (red).} 
\label{Fig088} 
\end{figure}

\begin{equation}
\delta V(t)=\frac{1}{K}\delta\nu(t)+\eta\delta P(t)\ ,\label{eq08N1}
\end{equation}
where $K$ is given by Eq. (\ref{eq01}) and $\eta$ is the response of the preamplifier detector, in V/W. $\delta P(t)$ holds for the power fluctuations and $\delta\nu(t)$ represents the frequency noise of the OPO. In Eq. (\ref{eq08N1}), the first term comes from frequency fluctuations and the second one from power fluctuations. We have also supposed in this equation that the fluctuations are small enough to allow us to keep only linear terms. In free-running mode, i. e. when the OPO frequency is not locked, comparison of the blue spectrum of Fig.\,\ref{Fig055}(b) with the green spectrum of Fig. \ref{Fig088} shows that the frequency noise term dominates over the intensity fluctuations: the second term of Eq.~(\ref{eq08N1}) can be neglected. However, comparison of the two spectra of Fig. \ref{Fig088} shows that this is no longer the case when the OPO is locked. If we suppose that the servo-lock is perfect and that the total voltage noise $\delta V(t)$ in Eq. (\ref{eq08N1}) is zero at the considered frequencies, Eq. (\ref{eq08N1}) gives a majorant to the frequency noise created at a given frequency $f$ by transfer from the intensity noise:
\begin{equation}
S_{\delta\nu}(f)=\frac{K^2}{\eta^2}S_{\delta P}(f)\ ,\label{eq08N2}
\end{equation}
where $S_{\delta P}(f)$ and $S_{\delta\nu}(f)$ are the power spectral densities of the power fluctuations and the frequency noise at frequency $f$, respectively. From Fig. \ref{Fig088}, we can see that the intensity noise is approximately 3 dB above the frequency noise. This means that the relative linewidth of Fig.\,\ref{Fig066}(b) is underestimated by a factor of approximately two. We can thus conclude that the linewidth of the locked OPO, with respect to the reference cavity, is of the order of 30\,Hz over 2\,s. This means that, even taking the detrimental effect of intensity noise into account, we have been able to decrease the linewidth of the non resonant idler by more than four orders of magnitude, well below the sub-kHz level.

It is worth noticing that such a low relative frequency noise of the OPO with respect to the reference cavity has been obtained without implementing a Pound-Drever-Hall stabilization scheme and with a relatively low finesse cavity \cite{Drever1983}. Of course, this linewidth is the relative linewidth with respect to the reference cavity, and any noise of the reference cavity will be transferred to the OPO idler, as we are going to investigate now.

\subsection{Long term stability}
\begin{figure}
\centering\includegraphics[width=0.9\columnwidth]{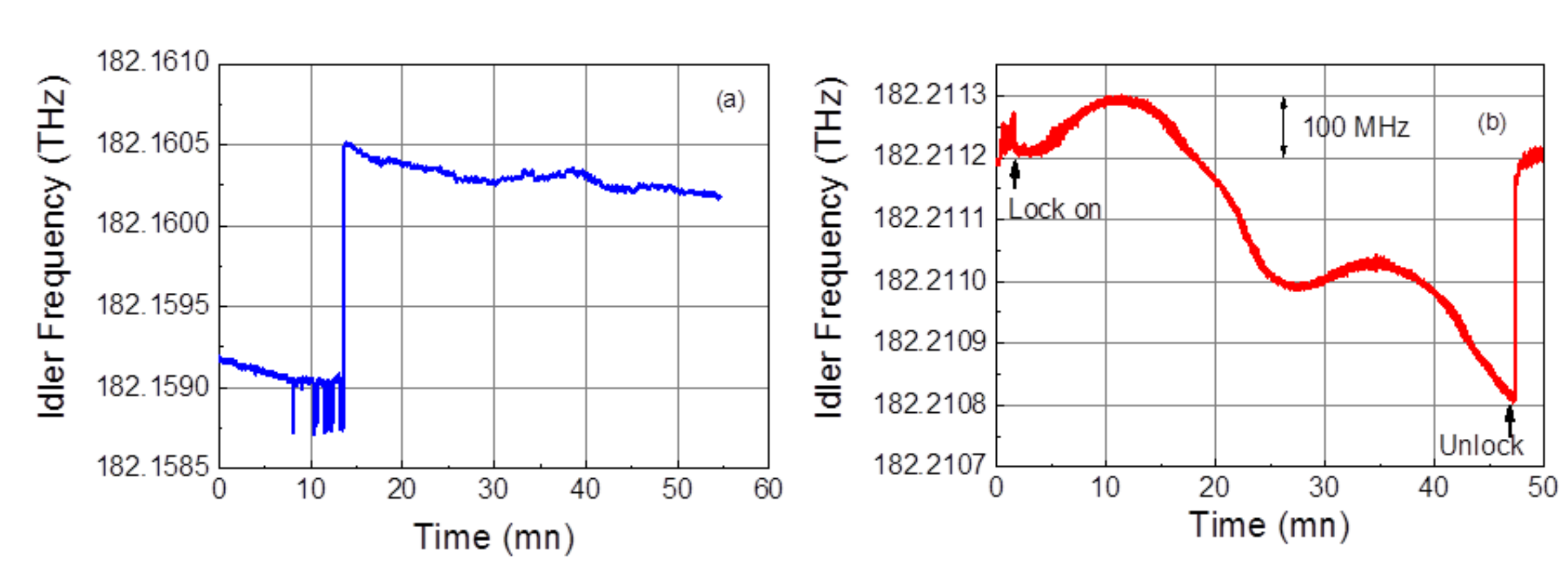}
 \vspace*{0cm} 
\caption{Long-term evolution of the non resonant idler frequency versus time. (a) Free running mode. (b) Locked mode.}
\label{Fig099} 
\end{figure}

The idler frequency long term stability is analyzed by using a 0.01~pm resolution wavelength meter (Angstrom WS7 from HighFinesse GmbH, see Fig.\ \ref{Fig011}). A typical result obtained when the OPO is free running (servo-locking OFF) is reproduced in Fig.\ \ref{Fig099}(a), for a 250-$\mu$m-thick \'etalon. This result shows that the free-running OPO typically experiences one mode-hop per hour. In the case reproduced in Fig.\ \ref{Fig099}(a), this mode hop corresponds to a few longitudinal modes. 

Once locked to the cavity, the idler frequency should become much more stable. This is what is observed in Fig.\ \ref{Fig099}(b). The OPO remains locked for 45~min, without any mode hop. During this time, the idler frequency drifts by 500\,MHz. This is due to the fact that the FP reference cavity, which is not thermally stabilized, is of course not an absolute frequency reference. This 500~MHz frequency excursion corresponds to a temperature variation of a fraction of a degree, which is perfectly plausible in our lab environment.

\begin{figure}
\centering\includegraphics[width=1.0\columnwidth]{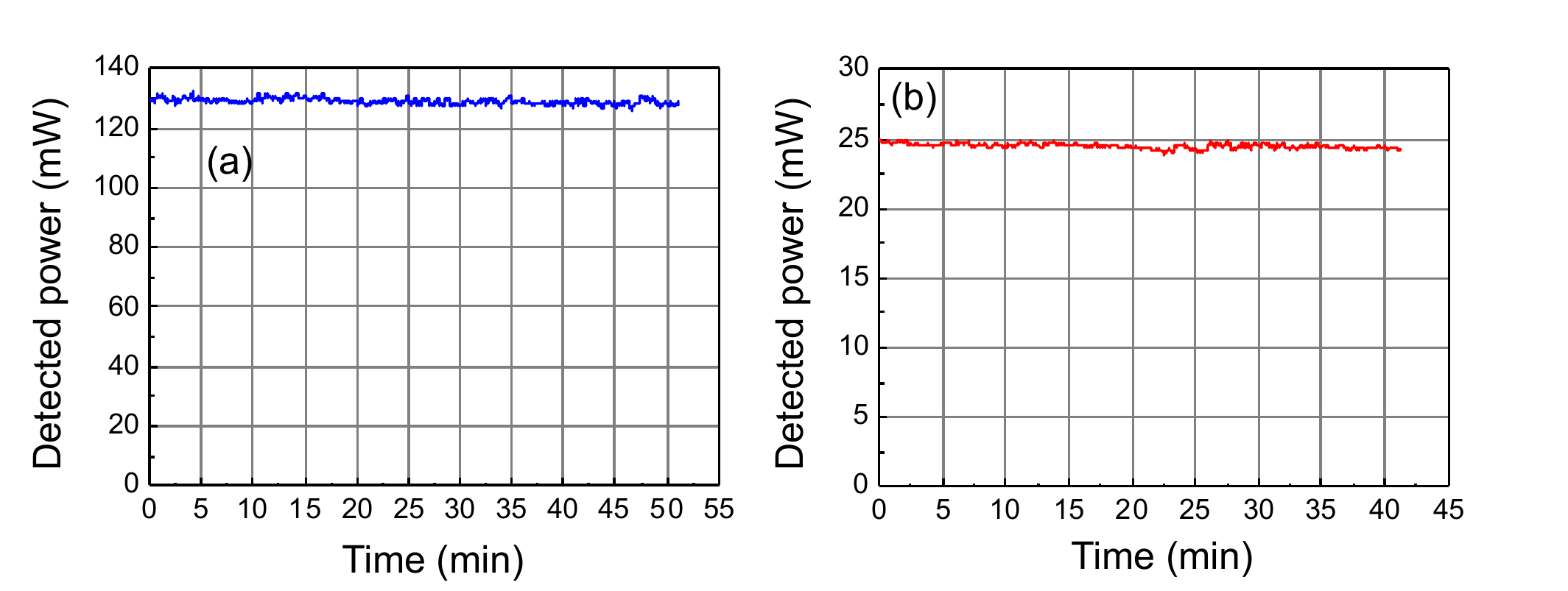}
 \vspace*{0cm} 
\caption{Time evolution of the non resonant idler power in (a) free running mode and (b) frequency locked regime.}
\label{Fig1010} 
\end{figure}

We also checked the power stability of the emitted idler, both in free running and frequency locked conditions (see Fig.~\ref{Fig1010}). The relative RMS power fluctuations are equal to 0.8~\%  in both the free running and locked modes. The frequency stabilization does not alter power stability, as expected.

\section{Conclusion}
\label{Conclusion}
In conclusion, we have demonstrated a cw SRO with output power of the order of 1 watt in the 1600--1700~nm wavelength domain and locked to the side of the transmission peak of a Fabry-Perot reference cavity. The relative frequency noise of the OPO with respect to the reference cavity has been shown to correspond to sub-kHz relative frequency deviations. With its low frequency noise, measured with respect to the reference cavity, combined to its good long term stability (frequency and power), this OPO source could be a good candidate to pump quantum interfaces necessary for quantum communication applications. Future development will include the thermal stabilization of the overall setup, and in particular the reference Fabry-Perot cavity. Better long term frequency stabilization can also be expected thanks to the use of a volume Bragg grating to replace one cavity mirror\cite{Zeil2014}. Finally, short term frequency fluctuations could also be reduced, together with immunity to intensity noises, by implementing a Pound-Drever-Hall locking scheme \cite{Drever1983} with a higher finesse cavity \cite{Mhibik2011}, although this is probably not necessary for our quantum interface application.

\section*{A. Derivation of the frequency noise fluctuations}
For a symmetric lossless FP cavity like that one used in our experiment, the transmission in the vicinity of a resonance frequency $\nu_0$ is given by:
\begin{equation}
T(\nu)=\frac{T_0}{1+\left(\frac{2F}{\pi}\right)^2\sin^2\left[\frac{\pi (\nu -\nu_0)}{\Delta}\right]}\ ,\label{eq08N3}
\end{equation}
where $T_0$ is the cavity transmission at resonance. If we call $\nu_1= \nu_0-\frac{\Delta}{2F}$ the locking point on the flank of the resonance peak, we can linearize Eq.\,(\ref{eq08N3}) around $\nu_1$  :

\begin{equation}
T(\nu(t)=\nu_1 +\delta\nu(t)) \approx T(\nu_1)+\delta\nu(t)\left.\frac{dT}{d\nu}\right|_{\nu=\nu_1}\ ,\label{eq08N4}
\end{equation}
where $\delta \nu (t)$  stands for the frequency fluctuations of the non resonant OPO idler which are supposed to be small compared to $\Delta/F$.  Using $T(\nu_1 )=T_0/2$  and $dT/d\nu(\nu_1)=FT_0/\Delta$, Eq. (\ref{eq08N4}) leads to:
\begin{equation}
\delta\nu(t) =\frac{\Delta}{F} \left(\frac{T\left[\nu (t)\right]-T_0/2}{T_0}\right)=\frac{\Delta}{F} \frac{\delta V (t)}{V_{PP}}\ ,\label{eq08N5}
\end{equation}
which leads to Eq.\,(\ref{eq01}).

This latter expression is valid under the following assumptions: i) The peak to peak amplitude of the frequency fluctuations should be small compared with $\Delta/F$ and ii) the frequency noise bandwidth should be less than $\Delta/2F$. The violation of one of these assumptions leads to an underestimation of the frequency noise.

\section*{Acknowledgment}
The authors are happy to thank S.~Tanzilli and F.~Kaiser for helpful discussions, and L.~Morvan for technical assistance. The authors also wish to thank one of the anonymous reviewers for having pointed out a mistake in the initial submission.



\end{document}